\documentclass[12pt]{article}
\usepackage{pdproc,epsfig}

\begin{document}

\renewcommand{\thefootnote}{\alph{footnote}}
\def\alt{\raise0.3ex\hbox{$\;<$\kern-0.75em\raise-1.1ex\hbox{$\sim\;$}}}
\def\agt{\raise0.3ex\hbox{$\;>$\kern-0.75em\raise-1.1ex\hbox{$\sim\;$}}}
\def\d{{\rm d}}
\newcommand{\be}{\begin{equation}}
\newcommand{\ee}{\end{equation}}
\newcommand{\bea}{\begin{eqnarray}}
\newcommand{\eea}{\end{eqnarray}}
\newcommand{\vv}{\,\,\, ,}
\newcommand{\pp}{\,\,\, .}

\title{ULTRAHIGH ENERGY NEUTRINOS WITH A MEDITERRANEAN NEUTRINO
TELESCOPE}

\author{E. Borriello, G. Miele and O. Pisanti}

\address{Dipartimento di Scienze Fisiche, Universit\'a di Napoli "Federico II" e INFN Sezione di Napoli,
Complesso Universitario di Monte S. Angelo, Via Cintia, Napoli,
80126, Italy}

\abstract{A study of the ultra high energy neutrino detection
performances of a km$^3$ Neutrino Telescope sitting at the three
proposed sites for ``ANTARES", ``NEMO" and ``NESTOR" in the
Mediterranean sea is here performed. The detected charged leptons
energy spectra, entangled with their arrival directions, provide
an unique tool to both determine the neutrino flux and the
neutrino-nucleon cross section.}

\normalsize\baselineskip=15pt
$\,\,\,$\\

Neutrinos are one of the main components of the cosmic radiation
in the ultra-high energy (UHE) regime.  Although their fluxes are
uncertain and depend on the production mechanism, their detection
can provide information on the sources and origin of the UHE
cosmic rays.

From the experimental point of view the detection perspectives are
stimulated by the several proposals and R\&D projects for Neutrino
Telescopes (NT's) in the deep water of the Mediterranean sea,
namely \verb"ANTARES" \cite{[16]}, \verb"NESTOR" \cite{[17]} and
\verb"NEMO" \cite{[18]}, which in the future could lead to the
construction of a km$^3$ telescope as pursued by the \verb"KM3NeT"
project \cite{[19],[20]}. Actually, on the \verb"ANTARES" site, a
smaller telescope with a surface area of 0.1 km$^2$ is already
under construction \cite{Aguilar:2006rm}. A further project is
\verb"IceCube", a cubic-kilometer under-ice neutrino detector
\cite{[21],[22],[23]}, currently being deployed in a location near
the geographic South Pole in Antarctica. \verb"IceCube" applies
and improves the successful technique of \verb"AMANDA" to a larger
volume.

\begin{figure}[p]
\begin{center}
\hspace{2cm}
\includegraphics[width=.70\textwidth]{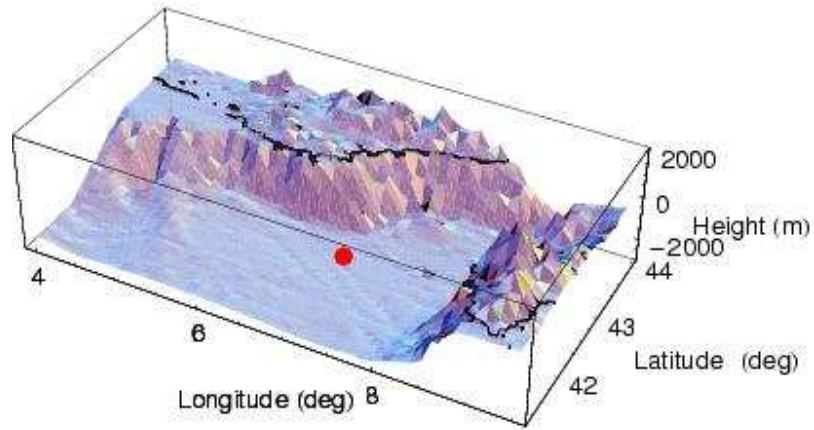}
\vspace{-0.0cm} \caption{The surface profile of the area near the
\texttt{ANTARES} site (red spot) at 42$^\circ$ 30' N, 07$^\circ$
00' E. The black curve represents the coast line. The sea plateau
depth in the simulation is assumed to be 2685 m. The effective
volume starts at an height of 100 m from the seabed, to account
for the spacing of the first photomultipliers as foresee by the
current designs.} \label{Antares}
\end{center}
\end{figure}
\begin{figure}[p]
\begin{center}
\hspace{2cm}
\includegraphics[width=.70\textwidth]{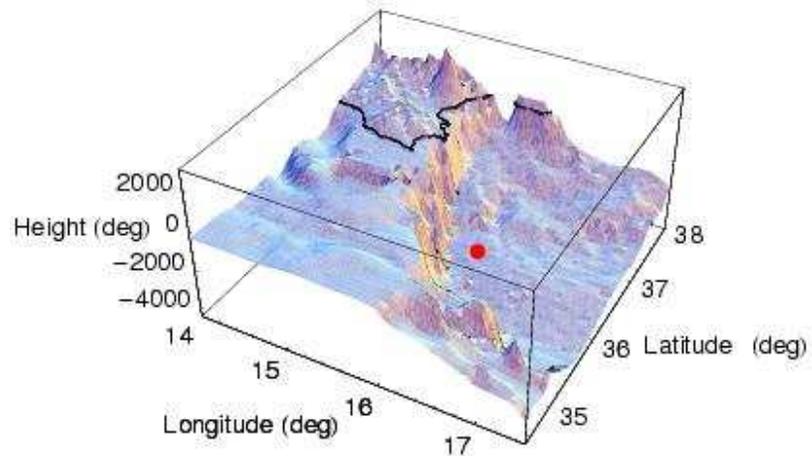}
\vspace{-0.8cm} \caption{The surface profile of the area near the
\texttt{NEMO} site (red spot) at 36$^\circ$ 21' N, 16$^\circ$ 10'
E. The black curve represents the coast line. The sea plateau
depth used in the simulation is 3424 m. The effective volume
starts at an height of 100 m from the seabed, to account for the
spacing of the first photomultipliers as foresee by the current
designs.} \label{Nemo}
\end{center}
\end{figure}
Although NT's were originally thought as $\nu_{\mu}$ detectors,
their capability as $\nu_{\tau}$ detectors has become a hot topic
\cite{Gandhi:1995tf,[24],[25],Anchordoqui:2005is,Yoshida:2003js,Beacom:2003nh,Athar:2000rx,Bugaev:2003sw,Ishihara},
in view of the fact that flavor neutrino oscillations lead to
nearly equal astrophysical fluxes for the three neutrino flavors.
Despite the different behavior of the produced tau leptons with
respect to muons in terms of energy loss and decay length, both
$\nu_\mu$ and $\nu_{\tau}$ detection are sensitive to the matter
distribution near the NT site. Thus, a computation of the event
detection rate of a km$^3$ telescope requires a careful analysis
of the surroundings of the proposed site. The importance of the
elevation profile of the Earth surface around the detector was
already found of some relevance in Ref.\ \cite{Miele:2005bt},
where some of the present authors calculated the aperture of the
Pierre Auger Observatory \cite{Auger,Abraham:2004dt} for
Earth-skimming UHE $\nu_{\tau}$'s. Indeed, air shower experiments
can be used as NT's at energies$\agt 10^{18}\,$eV, a topic
recently reviewed in \cite{Zas:2005zz}. In Ref.\
\cite{Cuoco:2006qd} it is estimated the effective aperture for
$\nu_\tau$ and $\nu_\mu$ detection of a km$^3$ NT in the
Mediterranean sea placed at any of the three locations proposed by
the \verb"ANTARES", \verb"NEMO" and \verb"NESTOR" collaborations.
The characteristics of the three site surface profiles
\cite{ETOPO2} are compared by using the DEM of the different
areas.

In the present paper we further develop the approach of Ref.
\cite{Cuoco:2006qd} in order to apply the detection of UHE $\nu$
as a tool to simultaneously measure the UHE neutrino flux and the
$\nu$-N cross section in extreme kinematical regions.

\begin{figure}[t]
\begin{center}
\hspace{2cm}
\includegraphics[width=.70\textwidth]{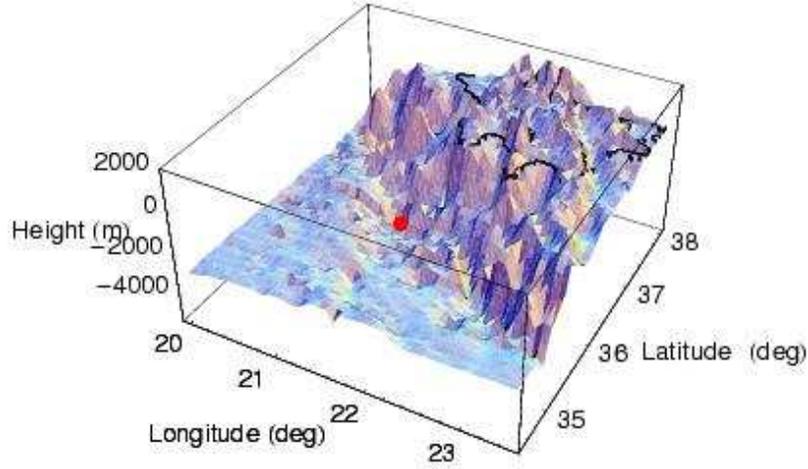}
\vspace{-0.8cm} \caption{The surface profile of the area near the
\texttt{NESTOR} site (red spot) at 36$^\circ$ 21' N, 21$^\circ$
21' E. The black curve represents the coast line. The sea plateau
depth in the simulation is assumed to be 4166 m. The effective
volume starts at an height of 100 m from the seabed, to account
for the spacing of the first photomultipliers as foresee by the
current designs.} \label{Nestor}
\end{center}
\end{figure}

\begin{figure}[t]
\begin{center}
\epsfig{figure=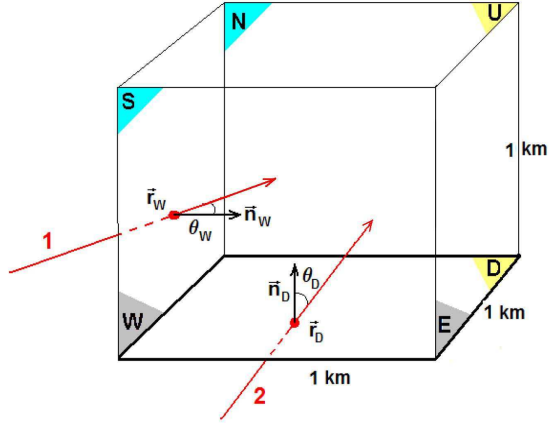,height=6cm} \caption{The angle
definition and the fiducial volume of a km$^3$ NT.} \label{kmcube}
\end{center}
\end{figure}

Following the formalism developed in \cite{Cuoco:2006qd} we define
the km$^3$ NT {\it fiducial} volume as that bounded by the six
lateral surfaces $\Sigma_a$ (the subindex $a$=D, U, S, N, W, and E
labels each surface through its orientation: Down, Up, South,
North, West, and East), and indicate with $\Omega_a \equiv
(\theta_a, \phi_a)$ the generic direction of a track entering the
surface $\Sigma_a$. The scheme of the NT fiducial volume and two
examples of incoming tracks are shown in Fig.\ \ref{kmcube}. We
introduce all relevant quantities with reference to $\nu_\tau$
events, the case of $\nu_\mu$ being completely analogous.

Let $\d \Phi_\nu/(\d E_\nu \, \d\Omega_a)$ be the differential
flux of UHE $\nu_\tau + \bar{\nu}_\tau$. The number per unit time
of $\tau$ leptons emerging from the Earth surface and entering the
NT through $\Sigma_a$ with energy $E_\tau$ is given by
\bea \left( \frac{\d N_\tau}{\d t} \right)_a = \int \d\Omega_a
\int \d S_a \int \d E_\nu \, \frac{\d\Phi_\nu(E_\nu, \Omega_a)}{\d
E_\nu\,\d\Omega_a} \int \d E_\tau \cos\left(\theta_a\right)
 k_a^\tau(E_\nu,E_\tau;\vec{r}_a,\Omega_a) \pp
\label{eq:1} \eea

The kernel $k_a^\tau(E_\nu\,,E_\tau\,;\vec{r}_a,\Omega_a)$ is the
probability that an incoming $\nu_\tau$ crossing the Earth, with
energy $E_\nu$ and direction $\Omega_a$, produces a $\tau$-lepton
which enters the NT fiducial volume through the lateral surface
$\d S_a$ at the position $\vec{r}_a$ with energy $E_\tau$ (see
Fig.\ \ref{kmcube} for the angle definition). If we split the
possible events between those with track intersecting the {\it
rock} and the ones only crossing {\it water}, the kernel
$k_a^\tau(E_\nu\,,E_\tau\,;\vec{r}_a,\Omega_a)$ is given by the
sum of these two mutually exclusive contributions,
\begin{equation}
k_a^\tau(E_\nu\,,E_\tau\,;\vec{r}_a,\Omega_a) =
k_a^{\tau,{r}}(E_\nu\,,E_\tau\,;\vec{r}_a,\Omega_a)+
k_a^{\tau,{w}}(E_\nu\,,E_\tau\,;\vec{r}_a,\Omega_a) \pp
\label{kern-split}
\end{equation}
For an isotropic flux we can rewrite Eq. (\ref{eq:1}), summing
over all the surfaces, as
\begin{eqnarray}
\frac{\d N_\tau^{(r,w)}}{\d t} = \sum_a \int \d E_\nu \, \int \d
E_\tau \int \d\Omega_a \int  \d S_a \left(
\frac{1}{4\pi}\,\frac{\d\Phi_\nu(E_\nu)}{\d E_\nu} \right)
\cos\left(\theta_a\right) \,
k_a^{\tau,{(r,w)}}(E_\nu\,,E_\tau\,;\vec{r}_a,\Omega_a)\pp\nonumber\\
\label{kernel1}
\end{eqnarray}
By using this expression one can also define the total aperture
$A^{\tau(r,w)}(E_\nu)$, with ``$r$" and ``$w$" denoting the {\it
rock} and {\it water} kind of events, respectively,
\begin{eqnarray}
\frac{\d N_\tau^{(r,w)}}{\d t} &=& \int \d E_\nu \,
\,\left(\frac{1}{4\pi}\,\frac{\d\Phi_\nu(E_\nu)}{\d E_\nu} \right)
\,A^{\tau(r,w)}(E_\nu)\vv \label{kernel2}
\end{eqnarray}
where
\begin{eqnarray}
A^{\tau(r,w)}(E_\nu) &=& \sum_a \int \d E_\tau \int \d\Omega_a
\int \d S_a \, \cos\left(\theta_a\right) \,
k_a^{\tau,{(r,w)}}(E_\nu\,,E_\tau\,;\vec{r}_a,\Omega_a) \pp
\label{kernel3}
\end{eqnarray}
Of course, the same quantities can be defined for muons coming
from the charged-current interactions of $\nu_\mu$.
\begin{figure}[t]
\begin{center}
\epsfig{figure=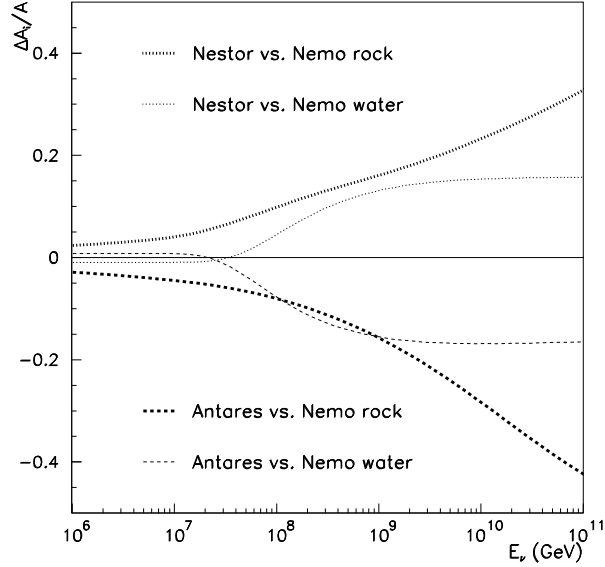,height=8cm} \caption{A comparison of
the effective apertures $A^{\tau((r,w))}(E_\nu)$ for the three NT
sites. We plot the ratios
$[A^{\tau((r,w))}(\mbox{\texttt{NESTOR}})-A^{\tau((r,w))}(\mbox{\texttt{NEMO}})]/A^{\tau((r,w))}(\mbox{\texttt{NEMO}})$
and
$[A^{\tau((r,w))}(\mbox{\texttt{ANTARES}})-A^{\tau((r,w))}(\mbox{\texttt{NEMO}})]/A^{\tau((r,w))}(\mbox{\texttt{NEMO}})$
versus the neutrino energy.} \label{comptelesc}
\end{center}
\end{figure}

In Fig.\ \ref{comptelesc} we compare the detection performances of
a km$^3$ NT placed at one of the three sites in the Mediterranean
sea. The \verb"NESTOR" site shows the highest values of the
$\tau$-aperture for both {\it rock} and {\it water}, due to its
larger depth and the particular matter distribution of the
surrounding area, while the lowest rates are obtained for
\verb"ANTARES". The aperture in the three sites can be quite
different at high energy but, in order to get the expected number
of UHE events per year, one has to convolve the aperture with a
neutrino flux which typically drops rapidly with the energy.
Although the percentage value of the matter effects remains
unchanged, in this very low statistics regime they can be hardly
distinguished; still, they can be enhanced by an appropriate
choice of the detector shape.
\begin{table}[b]
\centering
\begin{tabular}{|c|c|c|c|}
\hline Surf.
 & \verb"ANTARES" & \verb"NEMO" & \verb"NESTOR"\\
\hline
D & 0.0059/0 & 0.0059/0 & 0.0058/0 \\
U & 0/0.1677 & 0.0002/0.2133 & 0.0002/0.2543 \\
S & 0.0185/0.1602 & 0.0256/0.1773 & 0.0240/0.2011 \\
N &  0.0241/0.1540& 0.0229/0.1823 & 0.0321/0.1924 \\
W & 0.0212/0.1584 & 0.0335/0.1691 & 0.0265/0.2002 \\
E & 0.0206/0.1589 & 0.0190/0.1875 & 0.0348/0.1907 \\
\hline Total & 0.090/0.799 & 0.107/0.929 & 0.123/1.039\\
\hline
\end{tabular}
\caption{Estimated rate per year of {\it rock/water} $\tau$ events
at the three km$^3$ NT sites for a GZK-WB flux$^{23),25)}$. The
contribution of each detector surface to the total number of
events is also reported.} \label{table::events-WB}
\end{table}

Knowing the aperture of the NT at each site, we can compute the
expected $\tau$ event rate, once a neutrino flux is specified. In
Table \ref{table::events-WB} these rates are shown assuming a
GZK-WB flux \cite{Waxman:1998yy,Cuoco:2006qd}. The effect due to
the local matter distribution is responsible for the N-S, W-E and
NE-SW asymmetries for the \verb"ANTARES", \verb"NEMO" and
\verb"NESTOR" sites, respectively, as expected from the matter
profiles shown in Figs.\ \ref{Antares}, \ref{Nemo} and
\ref{Nestor}. These matter effects, for the specific UHE flux
considered (GZK-WB), correspond to an enhancement of {\it rock}
events which goes from 20 to 50\% for the three sites,
respectively, and a screening factor for {\it water} events from 3
to 10\%. The largest relative difference among lateral surfaces is
in the case of W/E for \verb"NEMO", where the huge wall to the
west of the site (see Fig. \ref{Nemo}) improves the rate by about
75\%, almost a factor 2! Notice also that the {\it water} events
from the U surface are basically proportional to the depth.

Due to the dependence of Eq. (\ref{kernel1}) on the neutrino flux
and the different behavior of
$k_a^{\tau,{r}}(E_\nu\,,E_\tau\,;\vec{r}_a,\Omega_a)$ and
$k_a^{\tau,{w}}(E_\nu\,,E_\tau\,;\vec{r}_a,\Omega_a)$ as functions
of the neutrino-nucleon cross section, $\sigma_{CC}^{\nu N}$, one
can imagine to use the detected events, properly binned for energy
loss and arrival direction, in order to obtain information on both
the neutrino flux and the neutrino-nucleon cross section. In
particular, since the real observable is the energy deposited in
the detector and not the energy and/or the nature of the charged
lepton, either $\mu$ or $\tau$, crossing the NT, one must sum the
two contributions. In fact, the events whose topology allows for
determining the nature of the charged lepton are a negligible
fraction of the expected total number.

\begin{figure}[t]
\begin{center}
\begin{tabular}{cc}
\includegraphics[width=.50\textwidth]{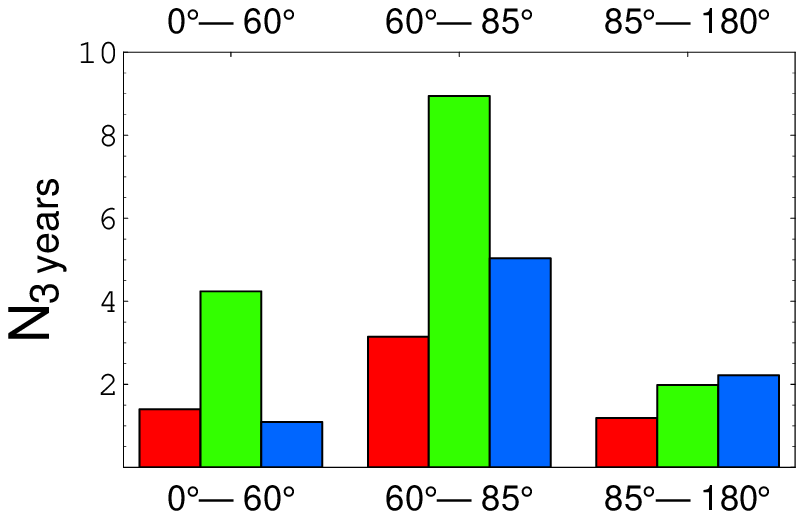} &
\includegraphics[width=.50\textwidth]{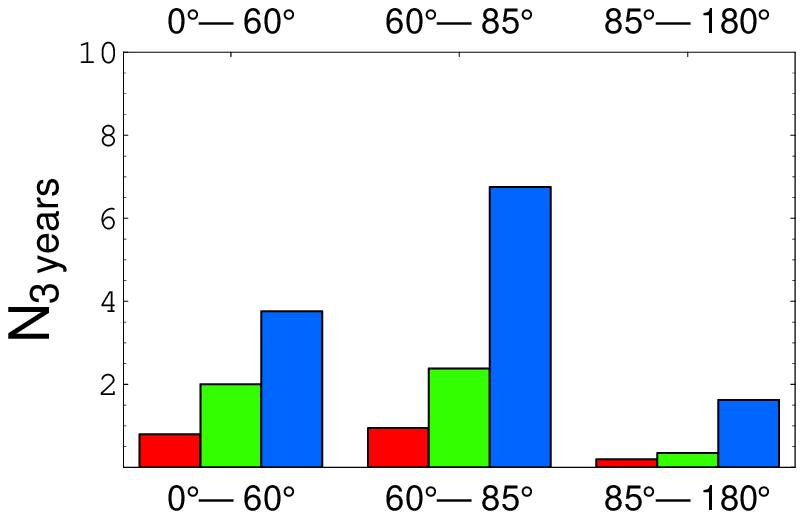}
\end{tabular}
\vspace{-0.8cm} \caption{In the two panels are reported the number
of events ($\mu+\tau$) collected in three years from a km$^3$ NT.
The left panel concerns the events with energy lost in the
detector in the range $10^5$-$10^8$ GeV, whereas on the right the
events have an energy deposited larger than $10^8$ GeV. See the
text for further details.} \label{bin1}
\end{center}
\end{figure}

\begin{figure}[p]
\begin{center}
\hspace{2cm}
\includegraphics[width=.50\textwidth]{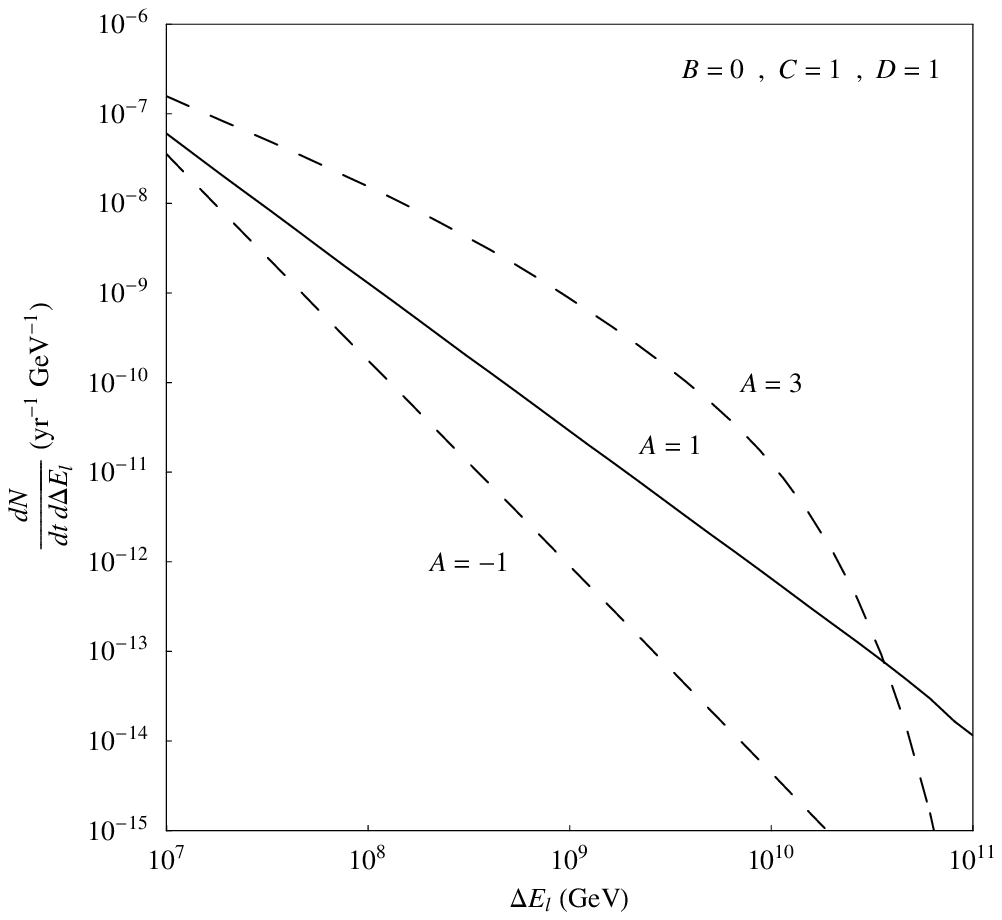}
\vspace{-0.8cm} \caption{The number density of yearly events
($\mu$ plus $\tau$) as function of the energy deposited by the
charged lepton $\Delta E_l$ for different values of $A$ and with
$B=0$ and $C=D=1$. The solid line represent the standard
scenario.}\label{PA}
\end{center}
\end{figure}
\begin{figure}[p]
\begin{center}
\hspace{2cm}
\includegraphics[width=.50\textwidth]{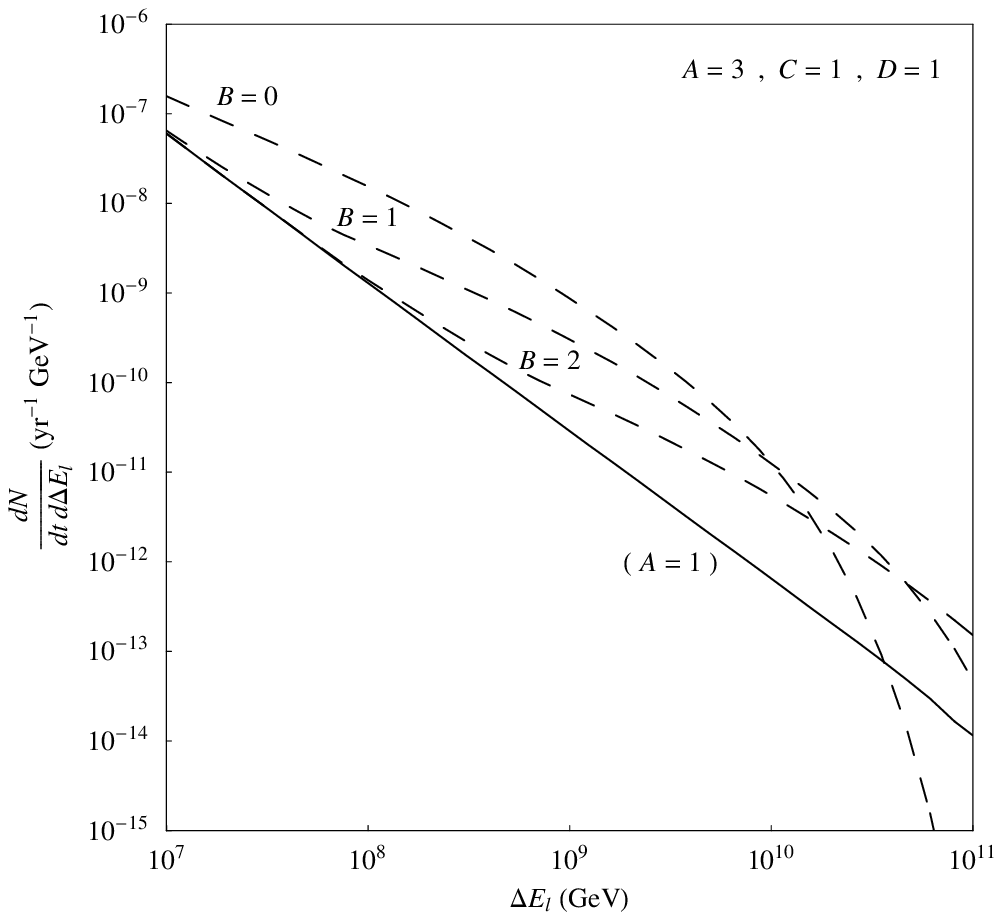}
\vspace{-0.8cm} \caption{The number density of yearly events
($\mu$ plus $\tau$) as function of the energy deposited by the
charged lepton $\Delta E_l$ for different values of $B$ and with
$A=3$ and $C=D=1$. The solid line represent the standard
scenario.} \label{PB}
\end{center}
\end{figure}
\begin{figure}[t]
\begin{center}
\hspace{2cm}
\includegraphics[width=.50\textwidth]{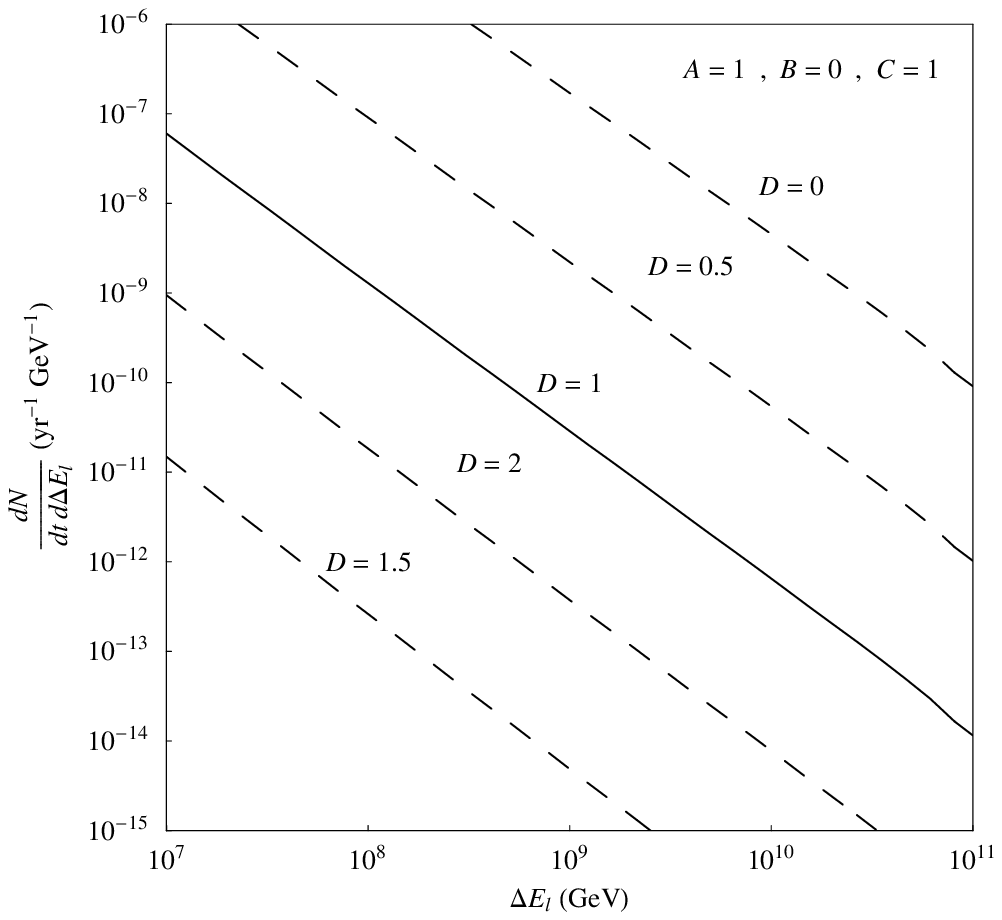}
\vspace{-0.8cm} \caption{The number density of yearly events
($\mu$ plus $\tau$) as function of the energy deposited by the
charged lepton $\Delta E_l$ for different values of $D$ and with
$A=C=1$ and $B=0$. The solid line represent the standard
scenario.} \label{PD}
\end{center}
\end{figure}

In the two panels of Figure \ref{bin1} are reported the number of
events ($\mu+\tau$) collected in three years from a km$^3$ NT. In
particular the left panel concerns the events where the energy
deposited in the detector is in the range $10^5$-$10^8$ GeV,
whereas the one on the right reports events where the energy lost
is larger than $10^8$ GeV. For each panel the events have been
split in three bins according to their arrival directions
($0^\circ$ represents the vertical downgoing direction). Fixing
the panel and the arrival direction range, the three bars of the
histogram represents three different neutrino fluxes and
$\sigma_{CC}^{\nu N}$ chosen. In particular from left to right we
have the GZK-WB$^{23),24)}$ flux and standard cross section,  the
GZK-WB$^{23),24)}$ flux and three times the standard cross section
and finally, the more copious GZK-H$^{23),24)}$ flux and standard
cross section. Note that the total number of events for GZK-WB
with $3 \, \sigma_{CC}^{\nu N}$ and for GZK-H with standard cross
section are the same. The different flux/cross section
configurations can be disentangled by observing the different
behavior of the height of the bars as function of the energy lost
and arrival direction.

In order to study the sensitivity  to both neutrino flux and
$\sigma_{CC}^{\nu N}$ it is necessary to parameterize their
standard expression and the possible departure from it. In
particular by using a standard Waxman-Bahcall
\cite{Anchordoqui:2005is} (C=D=1) as a conservative reference for
the neutrino flux we allow for a variation of its steepness via
the exponent $D$ and for the normalization through the
multiplicative factor $C$, namely:
\begin{equation}
 \phi_{WB} \cong 1.3 \cdot 10^{-8} \, C \,
\, \, \epsilon_\nu^{-2 \, D} \, \textrm{GeV$^{-1}$ cm$^{-2}$
s$^{-1}$ sr$^{-1}$} \pp
\end{equation}
In the same way, for the neutrino-nucleon cross section one can
parameterize the presence of new physics by assuming a departure
from the standard expression \cite{[27]} in terms of two free
parameters, $A$ and $B$, whose standard values are $A=1$ and
$B=0$:
\begin{eqnarray}
\sigma_{CC}^{\nu N} = 10^{-36} \, \textrm{cm}^{2}
\left\{\begin{array}{lr} 0.677 \, \epsilon_\nu^{0.492} \, ; & 2.00
\cdot 10^4 < \epsilon_\nu < 1.20 \cdot 10^7 \\ \\ 5.54 \,
\epsilon_\nu^{0.363} \, ; & 1.20 \cdot 10^7 < \epsilon_\nu < 1.20
\cdot
10^{7+B} \\ \\
5.54 \cdot 10^{0.363(1-A)(7.08+B)} \epsilon_\nu^{0.363 A} \, ; &
1.20 \cdot 10^{7+B} < \epsilon_\nu
\end{array}
\right.
\end{eqnarray}
where $\epsilon_\nu \equiv E_\nu/\textrm{GeV}$. In particular $B$
fixes the energy value where new physics appears and $A$ is the
change in the energy slope of $\sigma_{CC}^{\nu N}$. In Figures
\ref{PA}, \ref{PB} and \ref{PD} it is reported the effect of the
variation of the single parameter $A$, $B$, $D$ on the number
density of yearly events ($\mu$ plus $\tau$) as a function of the
energy deposited by the charged lepton $\Delta E_l$. The factor
$C$ has been fixed to its standard value ($C=1$) since it is just
a normalization and thus simply correlated to the exposure time
needed to achieve the proper event statistics.

The quite relevant effect shown by Figures \ref{PA}, \ref{PB} and
\ref{PD} supports once more the idea that a km$^3$ NT can provide
a real chance to both measure UHE neutrino flux and the
neutrino-nucleon cross section in the extreme kinematical region,
where some new physics could appear. Of course the real
feasibility of such measurements will crucially depend of the size
of the neutrino flux which fixes the time required to reach a
reasonable statistics. Then, while in this exercise we adopted an
extreme conservative point of view working with Waxman-Bahcall
like fluxes, one can wish for a more optimistic real situation.

\end{document}